\documentclass[pre,twocolumn,amsmath,amssymb,superscriptaddress]{revtex4}% Physical Review B
\usepackage{graphicx}
\usepackage{bm}
\usepackage{amsmath}
\usepackage{amsfonts}
\usepackage{amssymb}%
\usepackage{xcolor}

\newcommand{\be}{
\begin{equation}
}
\newcommand{\ee}{l
\end{equation}
}
\newcommand{\beq}{
\begin{eqnarray}
}
\newcommand{\eeq}{
\end{eqnarray}
}
\begin{document}
%\linenumbers
\title{ Inverse square L\'evy walks are not  optimal search strategies for $d\ge 2$\\
%{\small (A universal law of cell migration)}
 }

\author{Nicolas Levernier}
\affiliation{Department of Theoretical Physics, University of Geneva, Geneva, Switzerland}
\affiliation{Department of Biochemistry, University of Geneva, Geneva, Switzerland}

\author{Johannes Textor}
\affiliation{Institute for Computing and Information Sciences
Radboud University
Nijmegen, The Netherlands}

\author{Olivier B\'enichou}
\affiliation{Laboratoire de Physique Th\'eorique de la Mati\`ere Condens\'ee, UMR 7600 CNRS /UPMC, 4 Place Jussieu, 75255
Paris Cedex, France}

\author{Rapha\"el Voituriez}
\affiliation{Laboratoire de Physique Th\'eorique de la Mati\`ere Condens\'ee, UMR 7600 CNRS /UPMC, 4 Place Jussieu, 75255
Paris Cedex, France}
\affiliation{Laboratoire Jean Perrin, UMR 8237 CNRS /UPMC, 4 Place Jussieu, 75255
Paris Cedex}

\date{\today}

\begin{abstract}
The L\'evy hypothesis states that  inverse square L\'evy  walks are optimal search strategies because they maximise the encounter rate with sparse, randomly distributed, replenishable targets. It has served  as a theoretical basis to interpret a wealth of experimental data at various scales, from molecular motors to animals looking for resources, putting forward the conclusion that many living organisms  perform L\'evy  walks   to explore space because of their optimal efficiency.   Here  we  provide analytically the dependence on target density of the encounter rate of L\'evy walks for any space dimension $d$ ; in particular, this scaling is shown to be {\it independent} of the L\'evy exponent $\alpha$ for the biologically relevant case $d\ge 2$, which proves that the founding result of the L\'evy hypothesis
is incorrect. As a consequence, we  show that optimizing the encounter rate with respect to  $\alpha$ is {\it irrelevant} :  it does not change the scaling with density and can lead virtually to {\it any} optimal value of $\alpha$  depending on system dependent modeling choices. The conclusion that observed inverse square L\'evy patterns are the result of a common selection process based purely on the kinetics of the search behaviour is therefore unfounded.

\end{abstract}

\maketitle

\begin{figure}
\includegraphics[width=7cm]{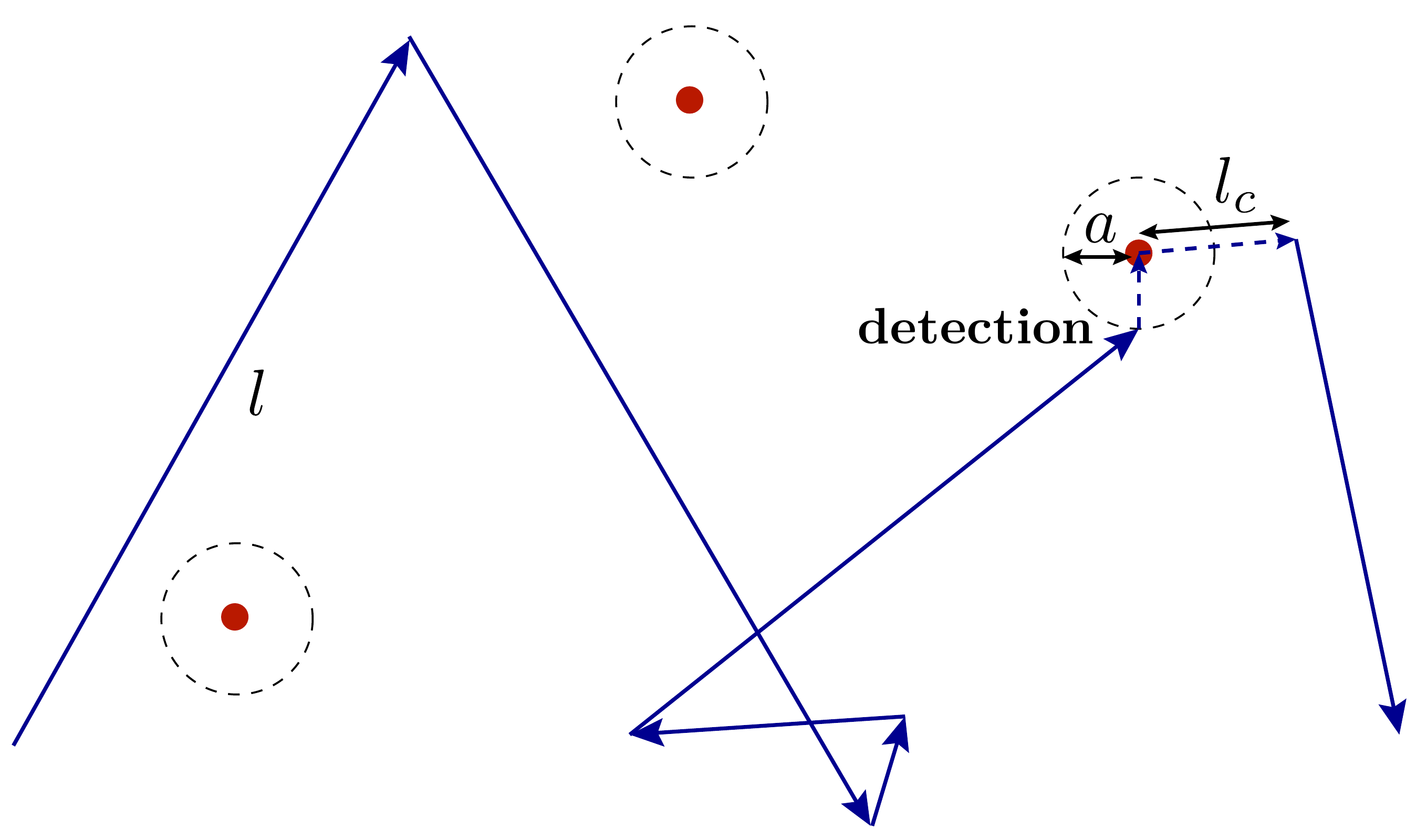}
\caption{ The L\'evy walk search model and its parameters. We consider a slightly more general version of the model originally    introduced in    \cite{Viswanathan:1999a}, here in $d=2$. The point-like searcher performs ballistic flights at constant speed $v$ (that can be set to 1) in uniformly distributed random directions. The length $l$ of each flight is drawn from a distribution   which satisfies  $p(l)\sim Cs^\alpha/l^{1+\alpha}$. In numerical simulations we used $p(l)=(2\pi)^{-1} l\int e^{ikl\cos\theta} e^{-s^\alpha k^\alpha}kdk d\theta$ for $d=2$. Targets are immobile and uniformly distributed in  infinite space with density $\rho$. A target is captured as soon as located within the detection radius $a$ (that can be set to 1) of the searcher, and is regenerated immediately after detection. To avoid systematic re-capture of the same target, an arbitrary rule is required, such as a cut-off time $\tau_c>a/v$ before target regeneration, or a cut-off distance from the target $l_c>a$ (which is the prescription that we used in numerical simulations) from which the walk is restarted. Finally, the model in its minimal form involves the following parameters : $a$ (that defines the unit length), $v$ (that defines the unit time), the target density $\rho$, the L\'evy exponent $\alpha$ and scale $s$ necessary to define $p(l)$, and the cut-off length $l_c$ (or equivalently a cut-off time $\tau_c$).}
\label{fig1}
\end{figure}

L\'evy walks \cite{Zaburdaev:2015aa} were introduced as a minimal random walk model that displays a superdiffusive scaling, while preserving a finite speed, and were originally motivated by various physical processes such as phase diffusion in Josephson junctions \cite{Geisel:1985aa,Shlesinger:1985aa} or passive diffusion in turbulent flow fields \cite{Shlesinger:1987aa}.
Shlesinger and Klafter   \cite{shles_klaft} were the first to report that, due to their weak oversampling properties, L\'evy walks provide a  more efficient way to explore space than normal random walks. This observation led Viswanathan et al.   \cite{Viswanathan:1996,Viswanathan:1999a}  to propose the following L\'evy search model (Fig.\ref{fig1}): they consider a searcher that performs ballistic flights of uniformly distributed random directions and constant speed, whose lengths $l$ are drawn from a distribution with power law tails $p(l)\sim Cs^\alpha/l^{1+\alpha}$  ($l\to\infty$) characterised by the L\'evy exponent $\alpha\in[0,2]$, where $s$ is a scale parameter and $C$ a dimensionless normalisation constant. The authors of    \cite{Viswanathan:1999a} consider an infinite space of dimension $d$ with Poisson   distributed (i-e with uniform density) immobile targets of density $\rho$, which are captured as soon as within a detection distance $a$ from the searcher. Two alternative hypotheses that   lead to two very different optimal strategies  ({\it i.e.}  strategies maximising the capture rate $\eta=\lim_{t\to\infty} n_t/t$ with respect to $\alpha$, where $n_t$ is the mean number of targets detected at  time $t$) are studied. (a) In the first case of  "revisitable targets", meaning that, as soon as detected, a target reappears and stays immobile at the same location, the authors claim that in the small density limit the encounter rate is optimized for a L\'evy exponent $\alpha\to 1$ ,   the so called inverse square L\'evy walk, and independently of the small scale characteristics of $p(l)$ or space dimension $d$. (b) In the second case of "destructive search" where each target can be found only once, the optimal strategy  is not of L\'evy type, but reduces to a simple linear ballistic motion for all $d$.

 The optimality of inverse square L\'evy walks claimed in    \cite{Viswanathan:1999a} is at the core of the L\'evy hypothesis, which 
 has been the reference theoretical framework for the analysis of trajectories of broad classes of living systems, from  molecular motors   \cite{Chen:2015tx} to cells   \cite{Harris:2012fk} and foraging animals     \cite{Miramontes:2014yo,Sims:2008,Humphries:2010,Viswanathan:1996,Viswanathan:1999a,hump2012} ; many studies  have indeed interpreted field data as L\'evy walks,   thereby concluding that their observation was the result of a selection process based on the optimality claimed in    \cite{Viswanathan:1999a}. In fact, since then the relevance to field data  of the condition (a) of revisitable targets  has been questioned  \cite{Benichou:2006lq,Benichou:2006,Tejedor:2012ly,Benichou:2011fk}, and  the identification of L\'evy patterns from real data has been debated    \cite{Edwards:2007,Benhamou:2007fk}. On the theoretical side, several alternative models, or variations of the original model  \cite{Viswanathan:1999a}  have been proposed \cite{Bartumeus:2016aa,Campos:2015aa,Bartumeus:2014aa,Wosniack:2017aa}. By allowing for more degrees of freedom, or by modifying the hypothesis of the original optimisation problem, these were shown to potentially lead to different optimal strategies. However, so far all studies  acknowledged the original result  \cite{Viswanathan:1999a}  as a founding benchmark in the field, and none has contested its technical validity.

In this letter,    we show on the basis of the same model that 
while the original analytical expression of the encounter rate with targets for L\'evy walks proposed in    \cite{Viswanathan:1999a} is correct in space dimension $d=1$,  it is incorrect for $d\ge 2$. As a consequence,   the conclusion that  inverse square L\'evy walks are  optimal search strategies  is not valid in the biologically relevant case  $d\ge 2$. In fact, relying on a recently developed framework to analyse non Markovian target search processes such as L\'evy walks
 \cite{Guerin:2016qf,Levernier:2018qf}, we show that,  as opposed to what is claimed in    \cite{Viswanathan:1999a}, 
for $d\ge 2$   the encounter rate  of L\'evy walks with sparse Poisson distributed targets (i) displays a linear dependence on  the concentration of targets {\it for all} values  of the L\'evy exponent, and  (ii)  can therefore be only marginally maximised, and  {\it for a broad range of values}  of the L\'evy exponent controlled by model dependent parameters, which makes the optimisation non universal. This  invalidates the claim that 
inverse square L\'evy walks  are optimal search strategies for $d\ge 2$, and more generally makes the optimisation of L\'evy search processes with respect to the L\'evy exponent non robust and thus irrelevant biologically for $d\ge2$. The conclusion   that observed inverse square L\'evy patterns across very different systems are the result of a common selection process based purely on the kinetics of the search behaviour is therefore unfounded.

Technically, it is straightforward to show  that for the case (b) of destructive search the optimal search strategy is achieved for $\alpha \to 0$ (straight ballistic motion), as stated in    \cite{Viswanathan:1999a} ; the ballistic strategy indeed minimises oversampling of space, as discussed in    \cite{shles_klaft}. This intuitive argument however fails in the case (a) of revisitable targets, which we discuss from now on. Let us first note that $1/\eta\equiv T$ is the mean time elapsed between successive capture events or in other words the mean first-passage time (MFPT) to any target for a searcher that starts immediately after a capture event. While the determination of  MFPTs of random walks has been studied at length in the literature because of the relevance of this observable to various fields 
   \cite{Redner:2001a,Shlesinger:2006,Condamin2007,Shlesinger:2007vf,bookSid2014}, its analytical calculation for non Markovian random walks, such as L\'evy walks, has remained until recently a technical challenge.  For that reason,  the analytical determination of $\eta$ proposed originally in     \cite{Viswanathan:1999a}  involved  uncontrolled hypothesis, and this result was proved correct analytically later in     \cite{Buldyrev:2001}, but only for $d=1$. Of note, in     \cite{Viswanathan:1999a} the predicted scaling of $\eta$  with target density 
was  supported  by  numerical simulations, but again  only for  $d=1$ ; in $d=2$, numerical simulations were shown for a single value of the density, thereby precluding any comparison with the predicted scaling.

\begin{figure}
\includegraphics[width=7cm]{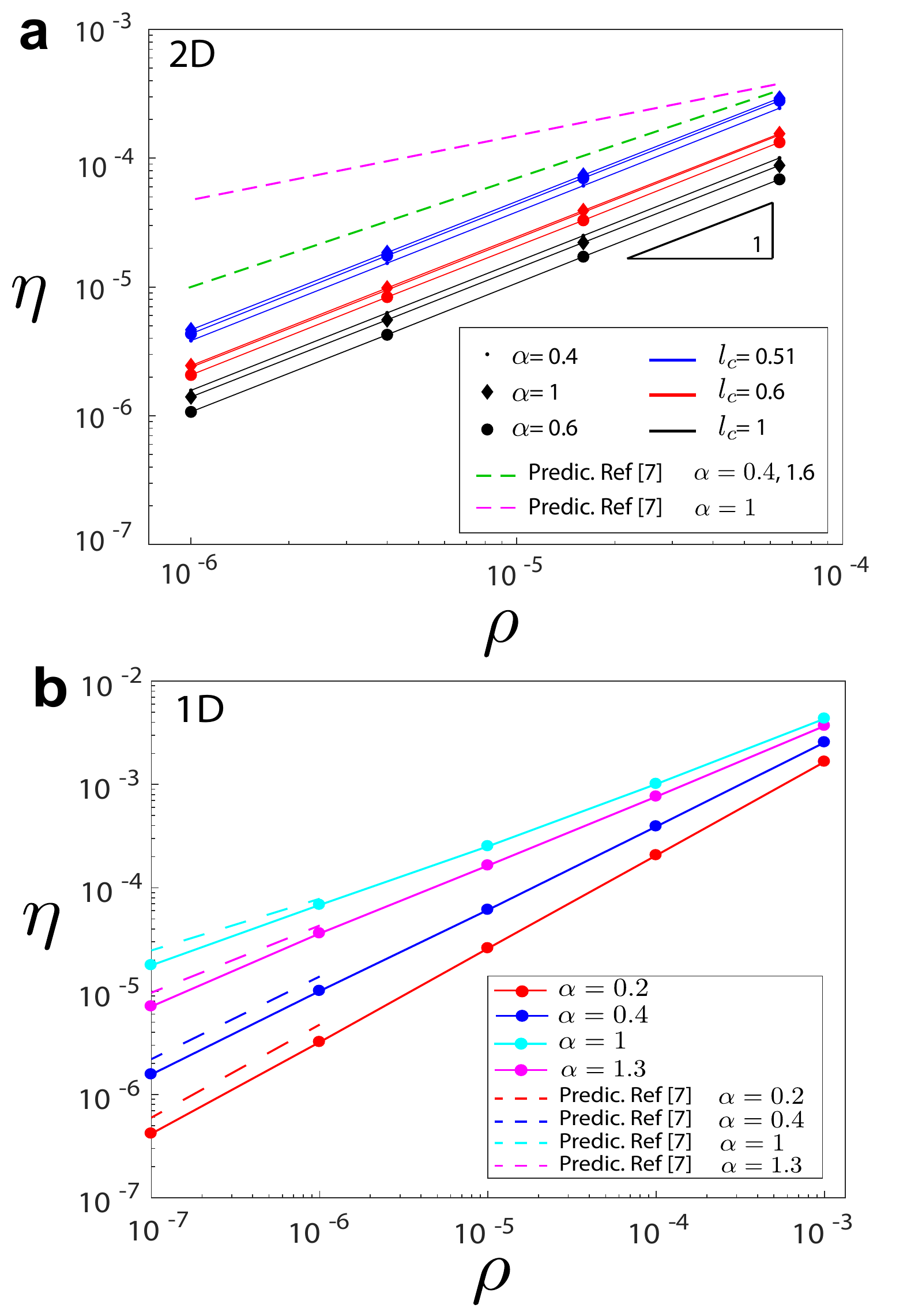}
\caption{The capture rate of L\'evy walkers has different scalings with  target density   for $d=2$ and $d=1$.  Capture rate $\eta$  as a function of target density ($\rho$) for different values of $\alpha$. {\bf a}. Case $d=2$.  Simulations are performed with 4000 Poisson-distributed targets of detection radius $a=0.5$   in a $2d$ square box of linear size $200/\sqrt{\rho}$ with periodic boundary conditions, following the dynamics defined in Fig.\ref{fig1} with $s=0.1$. Upon each detection event,  the searcher stops  and restarts immediately  from a distance $l_c$ from the  target.  In all cases, it is found that $\eta$ grows linearly with $\rho$ as predicted by Eq.\ref{Tmin}. Numerical simulations (symbols and plain lines) are compared to the predicted scaling of  \cite{Viswanathan:1999a} (dashed lines), and to our linear prediction (slope 1). {\bf b.}  Case $d=1$. Simulations are performed with Poisson distributed targets and make use of the dynamics defined in    \cite{Viswanathan:1999a}. The jump distribution is a truncated Pareto law : $p(l) = C/l^{1+\alpha}$ for $l>l_c$ and $p(l)=0$ for $l<l_c$, where $C$ is a normalisation constant; here $l_c=a=1$. Numerical simulations (symbols and plain lines) are compared to the predicted scaling of  \cite{Viswanathan:1999a} (dashed lines) that we recover in this paper.}
\label{fig2}
\end{figure}

Recently, new techniques have been introduced to determine analytically the MFPT of non-Markovian random walks to a single target in a confining volume $V$ in any space dimension $d$ in the large $V$ limit, first in the case of Gaussian processes    \cite{Guerin:2016qf}, and lately for general scale-invariant processes
   \cite{Levernier:2018qf}. Following a classical mean-field type argument    \cite{Berg:1976,Benichou:2011fk}, which was validated numerically, this result also yields in the large $V$ limit  the MFPT to  any target in infinite space with a concentration of Poisson distributed targets $\rho\equiv 1/V$, which is precisely the quantity that we aim at computing.  
   We here apply these techniques to L\'evy walks, which, importantly, requires to treat separately the cases of compact and non compact exploration.

    In the case of a compact walk, which occurs for L\'evy walks for $d=1$, it is found  \cite{Levernier:2018qf}  that the MFPT is given by
\begin{equation}
 T\underset{V \to \infty}{\sim} A\, V^{d_w(1-\theta)/d}\,{l_c}{^{d_w \theta}}
\end{equation}
where $A$ is a numerical constant, $d_w$ the walk dimension,  $\theta$ the persistence exponent, and $l_c$ the cut-off length introduced in Fig.\ref{fig1}. The walk dimension is given by $d_w=1$  for $\alpha<1$, and $d_w=\alpha$  for $\alpha>1$, and the persistence exponent is $\theta=\alpha/2$ for $\alpha<1$ and  $\theta=1/2$ for $\alpha>1$   \cite{Levernier:2018qf}. Thus, we get
\begin{equation}
\label{compact}
 T\underset{V \to \infty}{\sim}   \left \{
\begin{array}{ll}
D_1(\alpha)\, V^{1-\alpha/2} & \;\;(\alpha<1) \\
D_1(\alpha)\, V^{\alpha/2} & \;\;(\alpha>1) 
\end{array}
\right .
\end{equation}
where $D_1$ is  a numerical constant independent of the volume $V$.

   In the case of  non-compact random walks, which is the case of L\'evy Walks for $d\ge2$, the MFPT satisfies  \cite{Levernier:2018qf}
\begin{equation}
 T\underset{V \to \infty}{\sim} A\frac{V^{(d_w+\psi)/d}}{{a}^{\psi}} \left[1-B\left(\frac{a}{l_c}\right)^\psi \right] .
\end{equation}
In the latter, $A$ and $B$ are numerical constants, $d_w$ denotes the walk dimension and $\psi$ the transience exponent. Their corresponding values for a L\'evy walk of  parameter $\alpha$ is given by $d_w=1$ and $\psi=d-1$ for $\alpha<1$, and $d_w=\alpha$ and $\psi=d-\alpha$ for $\alpha>1$   \cite{Levernier:2018qf}. Hence, for any $\alpha \in [0,2]$, we get
\begin{equation}
 T\underset{V \to \infty}{\sim}D_{d}(\alpha, l_c/a)\; V  \hspace{1cm} 
\label{noncompact}
\end{equation}
where $D_{d}(\alpha, l_c/a)$ is a numerical constant depending on $\alpha$, $d$ and the microscopic parameters $l_c$ and $a$, but not on the  volume $V$.
\begin{figure}
\begin{center}
\includegraphics[width=7cm]{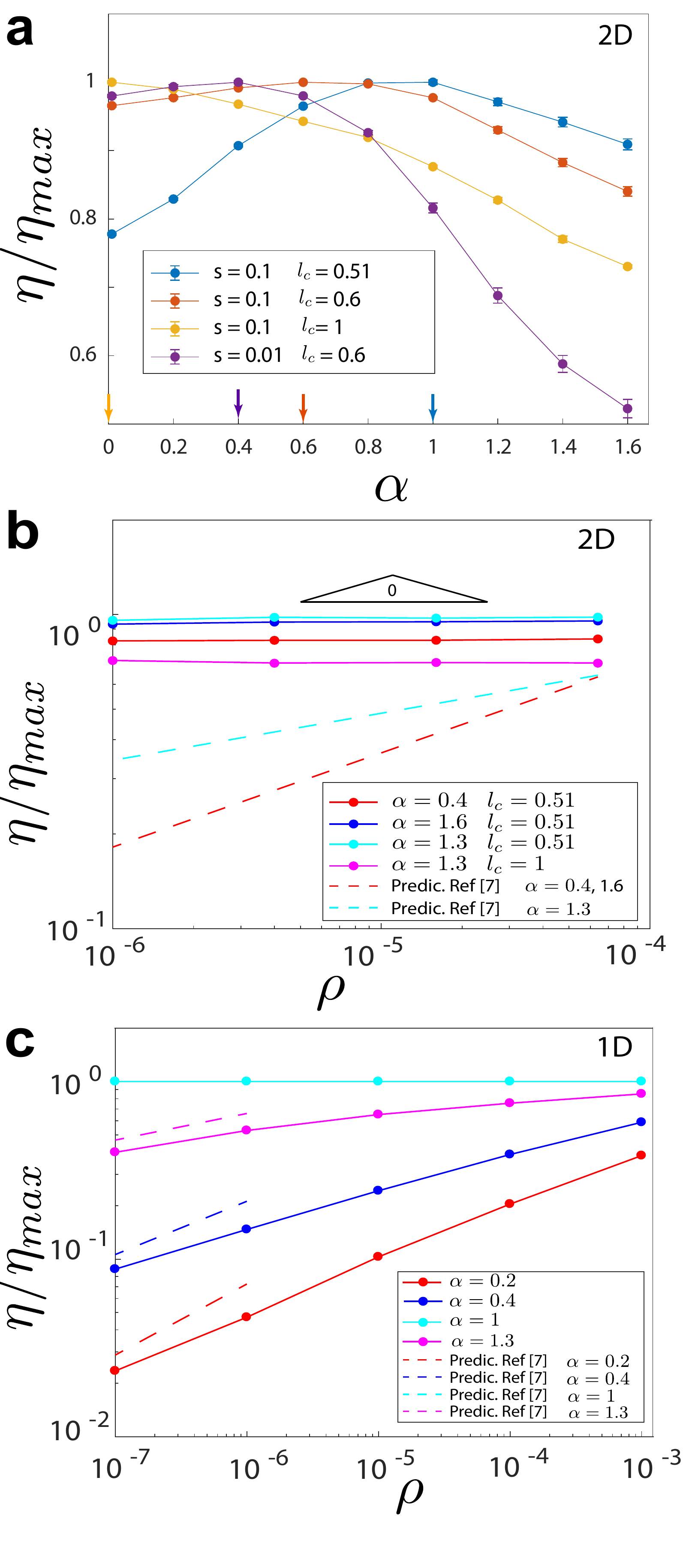}
\end{center}
\caption{A broad range of values  of the L\'evy exponent $\alpha$ can optimise the capture rate for $d=2$.
{\bf a.} Normalised capture rate as a function of $\alpha$ for different values of the cut-off distance $l_c$ and   scale parameter $s$ for $d=2$.  Simulations (symbols and lines) are performed with 4000 Poisson-distributed targets of detection radius $a=0.5$   in a $2d$ box of size $1000\times200$ with periodic boundary conditions.  The capture rate can be maximised for different values  $\alpha\in[0,2] $ (arrows)  depending on the  choice of parameters $l_c$ and $s$. {\bf b}.   Normalised capture rate as a function of $\rho$ for different values of $\alpha$ for $d=2$. Simulations (symbols and lines)  are performed with 4000 Poisson-distributed targets of detection radius $a=0.5$   in a $2d$ boxes of various sizes, with periodic boundary conditions and agree with our prediction (independence on $\rho$, slope 0).  Dashed lines show the  prediction of Ref.    \cite{Viswanathan:1999a}, which is invalid for $d=2$. The gain is bounded and independent of the target density $\rho$ for $\rho\to 0$, as predicted by Eq.\ref{Tmin}. {\bf c.}   Normalised capture rate as a function of $\rho$ for different values of $\alpha$ for $d=1$ (same dynamics as in Fig.\ref{fig2}). Simulations (symbols and lines)  are in agreement with the  prediction of  Eq.\ref{Tmin} (consistent with Ref.    \cite{Viswanathan:1999a}) shown in dashed lines. The gain diverges in the limit $\rho\to 0$. Note however the slow convergence to the exact scaling when $\rho\to 0$. }
\label{fig3}
\end{figure}

Using \eqref{noncompact} and \eqref{compact} with $\rho=1/V$, we finally   obtain analytically the mean capture rate  $\eta$,  thereby solving explicitly the original problem introduced in    \cite{Viswanathan:1999a} and recalled above  (Fig.\ref{fig1}). It is found that 
\begin{equation}
\label{Tmin}
\eta(\rho)\mathop{\sim}_{\rho\to 0}   \left \{
\begin{array}{ll}
K_1(\alpha)\rho^{1-\alpha/2}& d=1\ {\rm and} \ 0<\alpha<1  \\
K_1(\alpha)\rho^{\alpha/2}& d=1\ {\rm and} \ 1<\alpha<2  \\
K_d(\alpha)\rho& d\ge 2\ {\rm and} \ 0<\alpha<2  \\
\end{array}
\right .,
\end{equation}
where the constants $K_d$ are independent of $\rho$, but depend on the cut-off length $l_c$  that characterises the condition of restart after a capture event and the scale parameter $s$ that enters the definition of $p(l)$ (see Fig.\ref{fig1} ; this formulation contains in particular the original model as introduced in  \cite{Viswanathan:1999a} ). This result is valid asymptotically in the relevant  limit of sparse  targets  ($\rho\to 0$), and was checked numerically in Fig.\ref{fig2},\ref{fig3}.%Of note, the scaling of the capture rate with the density of target is controlled solely by the L\'evy exponent $\alpha$, and is independent of the small scale features of the model parametrised here by $l_c$ and $s$.

Several comments are in order. (i) For $d=1$, the result of Eq.\ref{Tmin} is consistent with the original result for $\eta(\rho)$ given in     \cite{Viswanathan:1999a}, as confirmed numerically in Fig.\ref{fig2}b and  analytically in      \cite{Buldyrev:2001}; in particular, in the $\rho\to 0$ limit,  $\eta$ is maximised  for the inverse square L\'evy walk $\alpha=1$, as claimed in     \cite{Viswanathan:1999a}. Of note, this optimum  is robust in the sense that the gain  $\eta_{\rm max}/\eta$ is arbitrarily large in the limit $\rho\to 0$ for all values of the  parameters $s$ and $l_c$, and is therefore critically controlled by the parameter $\alpha$ only (Fig.\ref{fig3}c). (ii) However, for the biologically relevant case $d\ge2$   the prediction for  $\eta(\rho)$ given in     \cite{Viswanathan:1999a}, claimed to be identical to the $d=1$ case, is incorrect. Indeed, the result of Eq.\ref{Tmin} shows that   $\eta$ depends linearly on  $\rho$  {\it for all}  $\alpha$ in contrast to the $d=1$ case  (confirmed numerically in Fig.\ref{fig2}a). (iii) This has strong consequences on the maximisation of $\eta$.  In fact, for $d\ge 2$, the dependence of $\eta$ on $\alpha$ lies only in the prefactor $K_d$. This implies first that the gain $\eta_{\rm max}/\eta$ achieved by varying $\alpha$ is bounded even in the limit $\rho\to 0$. In other words, tuning $\alpha$ can only yield a marginal gain, and therefore does not present a decisive selective advantage, as opposed to the $d=1$ case (Fig.\ref{fig3}a,b). Second, as we show numerically (Fig.\ref{fig3}a), $K_d(\alpha)$ presents bounded variations that depend drastically on the choice of parameters $s$ and $l_c$, which  could be arbitrary depending on the system studied. In particular, by performing minute variations of $s$ and $l_c$ it is found that $\eta$ can be maximised for {\it a broad range } of values of $\alpha\in[0,2]$, (Fig.\ref{fig3}). This overall makes the optimisation with respect to $\alpha$ biologically irrelevant for $d\ge 2$, and in particular invalidates the optimality of inverse square L\'evy walks claimed in \cite{Viswanathan:1999a} for generic values of $s$ and $l_c$.

These theoretical results have been fully validated by numerical simulations (Figs. \ref{fig2},\ref{fig3}), which confirm in particular the linear dependence of $\eta$ on $\rho$ independently of $\alpha$ for $d=2$ as predicted by Eq.(\ref{fig2}) (Fig. \ref{fig2}a), and the   sensitivity of $K_d$ to the system dependent  parameters $s$ and $l_c$  (Fig. \ref{fig3}a). In the context of animal foraging, the diverging gain at low target density obtained for $d=1$ (which could be relevant to specific biological examples) means that the implied optimal foraging behavior at $\alpha=1$ is expected to be a robust property that does not depend on the small scale characteristics of the specific biological system under study. Conversely, for $d\ge2$, which is the generic biologically relevant case, this conclusion does not hold because the optimal foraging behavior presents only a limited gain, and  may change even if seemingly minor changes are made to the system. For example,  very different optimal values of $\alpha$ can be obtained simply by allowing the searcher to have a short-term memory that would modify the small scale features of $p(l)$ or the way in which the first step after finding a target is performed (Fig.\ref{fig3}a). In fact, we found that $\alpha=1$ optimizes the encounter rate for $d=2$ only in the specific regime $l_c\to a$ and $s\ll a$, for which the problem is indeed expected to be effectively amenable to $d=1$ (see Fig. \ref{fig3}a).  Of note,  for $d\ge 2$ it is found numerically that $\eta$ seems to be always maximized for $\alpha < 2$, i--e  away from the Brownian limit, thereby suggesting that in this model L\'evy walks are more efficient than Brownian walks.  However, we stress again that the scaling of $\eta$ with target density is unchanged (up to logarithmic corrections for $d=2$), which makes the optimization overall of marginal importance.

Altogether, this shows that
inverse square L\'evy walks  are not generic optimal search strategies for $d\ge 2$, and therefore that the conclusion found in many studies  that observed inverse square L\'evy patterns are the result of a selection process is unfounded.   Importantly, we stress first that these results  do not invalidate the original   idea that L\'evy walks can be efficient to explore space    \cite{shles_klaft}, but disprove the specific role of inverse square L\'evy walks and their optimality for $d\ge 2$.  Second, on the experimental side, these results  do not question the validity of  observations of power law like patterns in field data, but refute the classical hypothesis that the observation of inverse square L\'evy walks would be the result of a selection process based  on the kinetics of the search behaviour.  Alternatively the observed patterns could be the result of various environmental parameters, such as the spatial distribution of prey, as suggested in    \cite{Humphries:2010} and observed in    \cite{Boyer:2006wk}.

{ \bf Acknowledgements}. This work was supported by  ERC grant FPTOpt-277998.

\end{document}